\UseRawInputEncoding
\documentclass[journal,twocolumn]{IEEEtran}
\usepackage{stfloats}
\usepackage{setspace}
\usepackage{algorithm}
\usepackage{algorithmic}
\usepackage{amsfonts}
\usepackage{dsfont}
\usepackage{enumerate}
\usepackage{cite,array}
\usepackage{graphicx}
\usepackage{lipsum}
\usepackage{amssymb}
\usepackage[cmex10]{amsmath}
\usepackage{cite}
\usepackage{mathrsfs}
\usepackage{graphicx}
\usepackage{float}
\usepackage{bm}
\usepackage{arydshln}
\usepackage{subfigure}
\usepackage{graphicx}
\usepackage{epstopdf}

\usepackage{color, soul, framed}
\usepackage[colorlinks,linkcolor=blue,citecolor=blue,urlcolor=blue]{hyperref}

\newtheorem{remark}{Remark}

\newtheorem{assumption}{Assumption}
\newtheorem{problem}{Problem}
\newtheorem{strategy}{Strategy}

\newcommand{\tr}{^{\mkern-1.5mu\mathsf{T}}}

\addtocounter{MaxMatrixCols}{20}

\begin{document}
\title{Distributed Eco-Driving Algorithm of Vehicle Platoon Using Traffic Light and Road Slope Information
\thanks{This research is partially supported by A*STAR under its RIE2020 Advanced Manufacturing and Engineering (AME) Industry Alignment Fund C Pre Positioning (IAF-PP) (Award A19D6a0053),
and partially supported by National Nature Science Foundation of China (Grant No. 62003218);
and partially supported by Guangdong Basic and Applied Basic Research Foundation (Grant No. 2019A1515110234).}}

\author{Yan Wang, Rong Su, Wei Wang, Xiaoxu Liu, Bohui Wang
\thanks{Y. Wang, R. Su and B. Wang are with the School of Electrical and Electronic Engineering,
Nanyang Technological University, 50 Nanyang Avenue, 639798, Singapore.  (E-mail:\small wang.yan@ntu.edu.sg;~rsu@ntu.edu.sg;~bhwang@ntu.edu.sg).}
\thanks{W. Wang is with the School of Management, Xi'an Jiaotong University, Xi'an 710049, China
(E-mail:\small wwwayne@xjtu.edu.cn).}
\thanks{X. Liu is with the Sino-German College of Intelligent Manufacturing, Shenzhen Technology University, Shenzhen, China (E-mail:liuxiaoxu@sztu.edu.cn).}
}
\maketitle


\IEEEpeerreviewmaketitle

\begin{abstract}
This paper investigates the problem of ecological driving (eco-driving) of vehicle platoons. To reduce the probability of the platoon avoiding red lights and increase fuel efficiency, a two-layer control architecture is proposed. The first layer is in charge of optimizing the leader's long-term motion profile using the traffic light and road slope information. The long-term planning model is defined based on the reachability analysis of the platoon to the green light windows. An event-triggered mechanism is proposed to operate the long-term planning model.
The second layer is the short-term adaptation, in which the leader attempts to follow the planning motion profile in real time, while the follower keeps track of the nearest preceding vehicle and the leader, to preserve the desired inter-vehicular distances.
A Newton's method-based algorithm is implemented to effectively solve both the long-term planning and short-term adaptation problems.
The effectiveness of the proposed formulation and algorithm is illustrated by the simulations.
\end{abstract}


\section{introduction}
Governments around the world have agreed to limit the gas emissions induced by the transportation \cite{european2011roadmap,undated2011epa}.
The emission rate of pollutant gases from a vehicle is positively correlated with the fuel consumption rate. The fuel consumption rate depends on many factors, including the vehicle characteristics,  road/traffic conditions, and driving behaviors.
With the development of sensor and communication technologies, the vehicle-to-vehicle(V2V) communication  and vehicle to infrastructure(V2I) communication are widely used in the field of transportation. For the vehicle platoon, the vehicles may obtain the road/traffic conditions and the motion planning of preceding vehicles in advance by V2I and V2V communications, respectively. The information obtained in advance can help design more reasonable driving strategies to reduce the fuel consumption rate.

The vehicle platoon may achieve a good performance by applying a proper control scheme \cite{wang2018decentralized,wang2020optimal}.
The application of PID/PID-like control strategies to vehicle platoons was studied in \cite{lui2020optimal,de20212,fiengo2019distributed}.
However, the fuel efficiency is not taken into account in \cite{wang2018decentralized,wang2020optimal,lui2020optimal,de20212,fiengo2019distributed}.
Considering the fuel efficiency, the eco-driving control strategies of vehicle/platoon have been designed from different perspectives in \cite{lim2016distance,yang2020eco,huang2018ecological,zhang2020optimized,turri2016cooperative,chen2018real,kamal2010board}. In particular, the eco-driving algorithm is designed based on the two-stage control architecture in \cite{lim2016distance,turri2016cooperative}. The authors of \cite{yang2020eco} studied the eco-driving control problem by a multi-objective optimization approach. The traffic state/environment is not considered in \cite{lim2016distance,yang2020eco}. To improve the practicality of the model, the traffic state/environment information is used for the vehicle  control strategy design in \cite{huang2018ecological,zhang2020optimized,turri2016cooperative,chen2018real,kamal2010board}.
However, the important traffic factor that is traffic light signal is not taken into account in
\cite{wang2018decentralized,wang2020optimal,lui2020optimal,de20212,fiengo2019distributed,lim2016distance,
yang2020eco,huang2018ecological,zhang2020optimized,turri2016cooperative,chen2018real,kamal2010board}. Traffic lights play an important role in the urban traffic. In this paper, we present a two-layer control architecture for the vehicle platoon eco-driving using traffic light and road slope information.

It has been shown that the reduction of red light idling can significantly improve the fuel efficiency for the vehicle \cite{homchaudhuri2016fast}.
Fuel economic vehicle control algorithm using traffic light signal obtained by V2I communication has received lots of research attention.
The reference \cite{asadi2010predictive} showed  how to use the upcoming traffic signal in the adaptive cruise control (ACC) system to reduce red light idling and fuel consumption.
The case of cooperative ACC for connected vehicles platoon is considered in  \cite{ma2021eco}.
%
%
For the multiple signalized intersections environment, a hierarchical control framework was proposed in \cite{du2017coordination} to coordinate a group of connected vehicles to reduce the red lights idling.  A 4-legged signalized intersection scenario is considered in \cite{zhang2019eco}, where  the eco-driving model with the cooperative vehicle-infrastructure systems is studied. The driver factors are not considered in \cite{asadi2010predictive,ma2021eco,du2017coordination,zhang2019eco}.
The literature \cite{zhao2018platoon,chen2021mixed} develops the eco-driving model under the mixed traffic conditions in which the human-driven vehicles are considered.
Considering the driver behavior and capability, a fuel economic driver assistant system¡¯s control strategy was developed in \cite{homchaudhuri2019control} for multiple connected vehicles under urban road conditions.
For the models proposed in \cite{homchaudhuri2016fast,asadi2010predictive,du2017coordination,homchaudhuri2019control,zhao2018platoon,chen2021mixed}, the traffic light information was described  as soft constraints used to reduce the red light idling.  The boundary condition methods and the speed guidance schemes are used in  \cite{ma2021eco} and \cite{zhang2019eco}, respectively, for reducing the red light idling.
In this paper, a reachability analysis-based position profile constraint is derived based on the traffic light signal for reducing the platoon red light idling.

The eco-driving control of platoon with dynamic topology is studied in \cite{zhang2019eco,yao2021lane,bevly2016lane}.
The platoon separation behavior is implemented by the speed guidance schemes calculated by a multi-objective optimization method in \cite{zhang2019eco}.
The lane-change behavior of the platoon is studied in \cite{yao2021lane}, which developed a decentralized lane-change-aware trajectory optimization model of platoon under the environment of signalized intersection with multi-lane roads. A survey of the lane change and merge maneuvers of platoon is reported in \cite{bevly2016lane}. The lane change and merge maneuvers are implemented by the longitudinal control and lateral control.
From the survey in \cite{bevly2016lane}, only a few studies have investigated the vehicle cut-in behavior by considering the fuel efficiency in the signalized intersections environment. The reason may be that the vehicle cut-in drastically increases traffic oscillation of the platoon and  the safety issue is the main concern for the  vehicle cut-in behavior.
The separate, lane change and merge behaviors of the platoon are usually implemented by the speed guidance schemes \cite{zhang2019eco,bevly2016lane}. How to design a precise speed guidance for ensuring safety, improving energy efficiency and avoiding red light is a challenging tradeoff problem.
In this paper, we only focus on the vehicle longitudinal control, and the change of the platoon topology is not considered.


 A distributed eco-driving problem of vehicle platoon is studied in this work. The leader receives the road environmental information from infrastructure by V2I communication. The followers receive the driving behavior prediction from its nearest preceding vehicle and the leader by V2V communication. A two-layer optimization architecture is proposed for the platoon to avoid red light and improve the fuel efficiency. For the first layer, the upcoming traffic light information and the slope of road ahead are used to define the long-term planning problem.
Firstly, the reachability of the platoon to the green light windows is analyzed. Based on the reachability analysis, the position constraints induced by the traffic light signal are established, and are used to define the long-term planning problem. An event-triggered mechanism is proposed to operate the long-term planning model.
For the second layer, motion predictions of the leader and the nearest preceding vehicle are applied to define the short-term adaptation problems.
An algorithm based on Newton's method is developed for solving both the long-term and short term problems.  The effectiveness of the proposed methods is illustrated by simulations.

The main contributions of this paper in comparison to the existing works are stated as follows: 1) A reachability analysis-based position profile constraint is developed for the platoon to avoid the red lights. Compared to the soft constraint method in \cite{homchaudhuri2016fast,asadi2010predictive,du2017coordination,homchaudhuri2019control,zhao2018platoon,chen2021mixed}, the platoon successfully avoids the red lights with a bigger probability by using our  method. 2) An event-triggered mechanism is developed to operate the long-term planning model such that the long-term planning data can be generated appropriately and timely.
3) A Newton¡¯s method-based algorithm is implemented to effectively solve the proposed new model within the acceptable time.



\section{Formulation}

\subsection{Vehicle Dynamic}
In this paper,
a platoon composed of heterogeneous vehicles is considered.
According to the Newton's second law, the longitudinal dynamics of a vehicle  can be described by:
\begin{align}
\label{eq1}
&v_{i}(k+1)=v_{i}(k)+a_{i}(k)\Delta t,\\
&s_{i}(k+1)=s_{i}(k)+v_{i}(k)\Delta t+\frac{1}{2}a_{i}(k)(\Delta t)^{2},\\
\label{a2}
&a_{i}(k)=\frac{1}{M_{i}}\Big(F_{i,T}(k)-F_{i,B}(k)-F_{i,E}(k)\Big),
\end{align}
where $v_{i}(k)$, $s_{i}(k)$ and $a_{i}(k)$ are the speed, position and acceleration of vehicle $i$ at time $k$; $M_{i}$ is the mass of vehicle $i$; $\Delta t$ is the sampling period;
$F_{i,T}(k)$ is the traction force; $F_{i,B}(k)$ is the brake force; and $F_{i,E}(k)$ is the resistance induced by the environment.
The environment resistance $F_{i,E}(k)$ is given by
\begin{align}
\label{ee4}
F_{i,E}(k)&=g\sin(\theta(s_{i}(k)))\nonumber\\
&\quad+gc_{i}^{r}\cos(\theta(s_{i}(k)))+F_{i,A}(k),
\end{align}
where the meanings of $g$, $\theta(s_{i}(k))$ and $c_{i}^{r}$ are given in Table I; the first term $g\sin(\theta(s_{i}(k)))$ is the force caused by gravity; the second term
$gc_{i}^{r}\cos(\theta(s_{i}(k)))$ is the rolling resistance; and the third term $F_{i,A}(k)$ is the air drag. It follows from \cite{homchaudhuri2016fast} that
\begin{align}
 F_{i,A}(k)=\xi_{i,d}(k)v_{i}^{2}(k),
\end{align}
where

\begin{align}
\label{zeq6}
\xi_{i,d}(k)&=\left\{
             \begin{array}{ll}
               \frac{1}{2}c_{d}\rho S_{i,A}, & i=1, \\
               \frac{1}{2}c_{d}\rho S_{i,A}\Big(1+\frac{\alpha d_{i,i-1}(k)-\beta}{100}\Big), & i>1,
             \end{array}
           \right.\\
\label{eq8}
d_{i,i-1}(k)&=s_{i-1}(k)-s_{i}(k),
\end{align}
where $\xi_{i,d}(k)$ is the air drag coefficient; the meanings of $c^{d},~\rho,~S_{i,A}$ are given in Table I.  Due to that
the vehicle $1$ ($i=1$) has no preceding vehicle, the air drag coefficient is of normal form $\xi_{1,d}(k)=\frac{1}{2}c_{d}\rho S_{1,A}$. It is known that the air drags of the follower vehicles will be reduced when the vehicles drive as a platoon.  For $i\geq2$, the variation of aerodynamics of vehicles should be taken in to account. Thus, the
air drag coefficient of vehicle $i$ ($i\geq 2$) is of a modified form $\xi_{i,d}(k)=\xi_{1,d}(k)(1+\frac{\alpha d_{i,i-1}(k)-\beta}{100})$ which depends on the inter-vehicular distance $s_{i-1}(k)-s_{i}(k)$.
\begin{remark}
The air drag on a vehicle is not only affected by its preceding vehicles,  but also by its follower vehicles.
The effect on the vehicle's air drag induced by its follower vehicle is smaller than the one induced by its preceding vehicle \cite{turri2016cooperative}. Thus, the effect of the follower vehicle on the vehicle's air drag is neglected in \cite{turri2016cooperative}.
Following \cite{turri2016cooperative}, we only consider the air drag reduction induced by the preceding vehicle in this paper.
\end{remark}

\begin{table}[h]
\centering
\caption{}
\label{tab:table}
\begin{tabular}{|c|c|}
\hline
$g$ & gravitational acceleration\\
$\theta(s)$ & the road slope at position $s$\\
$c_{i}^{r}$ & rolling resistance coefficient\\
$c^{d}$ & drag coefficient\\
$\rho$ & air density\\
$S_{i,A}$ & face area\\
\hline
\end{tabular}
\end{table}

\subsection{Fuel Consumption Rate}
For a typical vehicle, the expression of fuel consumption is complex and is unlikely to be accurately modeled. Most of the researchers  try to capture the fuel consumption model based on the regression of the raw data \cite{turri2016cooperative,kamal2012model,lim2016distance}.
Here, the fuel consumption rate is approximately modeled by the following polynomial:
\begin{align}
\label{q9}
P_{i}(k)=\begin{bmatrix}
           F_{i,T}(k) \\
           v_{i}(k) \\
         \end{bmatrix}\tr O_{i}\begin{bmatrix}
           F_{i,T}(k) \\
           v_{i}(k) \\
         \end{bmatrix}+\bar{O}_{i}\begin{bmatrix}
           F_{i,T}(k) \\
           v_{i}(k) \\
         \end{bmatrix}+\bar{o}_{i,3},
\end{align}
where $O_{i}=\begin{bmatrix}
               o_{i,11} & o_{i,12}\\
               o_{i,21} & o_{i,22} \\
             \end{bmatrix}
$, $\bar{O}_{i}=\begin{bmatrix}
        \bar{o}_{i,1} & \bar{o}_{i,2}\\
      \end{bmatrix}
$, $\bar{o}_{i,3}$ are the fitting coefficients.

\subsection{System Constraints}
For a typical vehicle, the traction force, brake force, and speed should be bounded, due to the physical constraints. A vehicle with dynamics \eqref{eq1}--\eqref{eq8} should satisfy:
\begin{align}
\label{zeq10}
 0\leq F_{i,T}(k)\leq \bar{F}_{i,T},\\
 0\leq F_{i,B}(k)\leq \bar{F}_{i,B},\\
\label{zz12}
0\leq v_{i}(k)\leq v_{\max}.
\end{align}


\subsection{Road Speed Limits}
In general, the vehicle on the road should satisfy the road speed limits:
\begin{align}
\label{b15}
v_{\min}\leq v_{i}(k)\leq v_{\max},
\end{align}
most of the time.
In this paper, we deal with $ v_{i}(k)\leq v_{\max}$ as a strict constraint, and the one $v_{i}(k)\geq v_{\min}$ as a soft constraint to enable the vehicle to run at a low speed in some special situations (for example, when vehicle starting up or slowing to a stop). However,
the platoon is not allowed to travel at a low speed (inferior to $v_{\min}$) as to avoid taking a red light.
This will be further discussed in subsections \ref{la}.

\subsection{Traffic Light Signal  Model}
The position of traffic light $j$ is $p_{j}$. Let $t_{j}^{r}$ and $t_{j}^{g}$ denote the red and green light durations of traffic light $j$, respectively. We use the cycling clock signal to indicate the state of the traffic light. The clock signal period of traffic light $j$ is $c_{j}=t_{j}^{r}+t_{j}^{g}$. The cycling
clock time of traffic light $j$ at time $k$ is denoted by $\tau_{j}(k)$ whose dynamic is
\begin{align}
 \tau_{j}(k+1)=(\tau_{j}(k)+1) \textmd{ mod } c_{j},
\end{align}
where $\tau_{j}(t)\in \{0,1,\cdots,c_{j}\}$. The state of the $j$-th traffic light is defined as
\begin{align}
\label{b16}
x_{j}(k)=\left\{
        \begin{array}{ll}
          1, & \textmd{ if } \tau_{j}(k)\in[0,t_{j}^{r}],\\
          0, & \textmd{ if } \tau_{j}(k)\in(t_{j}^{r},c_{j}],
        \end{array}
      \right.
\end{align}
where $x_{j}(k)=1$ and $x_{j}(k)=0$ mean that the traffic light $j$ is red and green at time $k$, respectively. From \eqref{b16}, the two upcoming green light windows are:
\begin{itemize}
  \item If $x_{j}(k)=1$, then
\begin{align}
\label{z17b}
&\quad[k+t_{j}^{r}-\tau_{j}(k), k-\tau_{j}(k)+c_{j}], \nonumber\\
&\textmd{ and }[k+t_{j}^{r}-\tau_{j}(k)+c_{j}, k-\tau_{j}(k)+2c_{j}].
\end{align}
  \item If $x_{j}(k)=0$, then
\begin{align}
\label{z18b}
&\quad[k, k-\tau_{j}(k)+c_{j}]\nonumber\\
&\textmd{ and } [k+t_{j}^{r}-\tau_{j}(k)+c_{j}, k-\tau_{j}(k)+2c_{j}].
\end{align}
\end{itemize}

\subsection{Eco-driving Framework}
\begin{figure}[!h]
  \centering
  \includegraphics[width=1\linewidth]{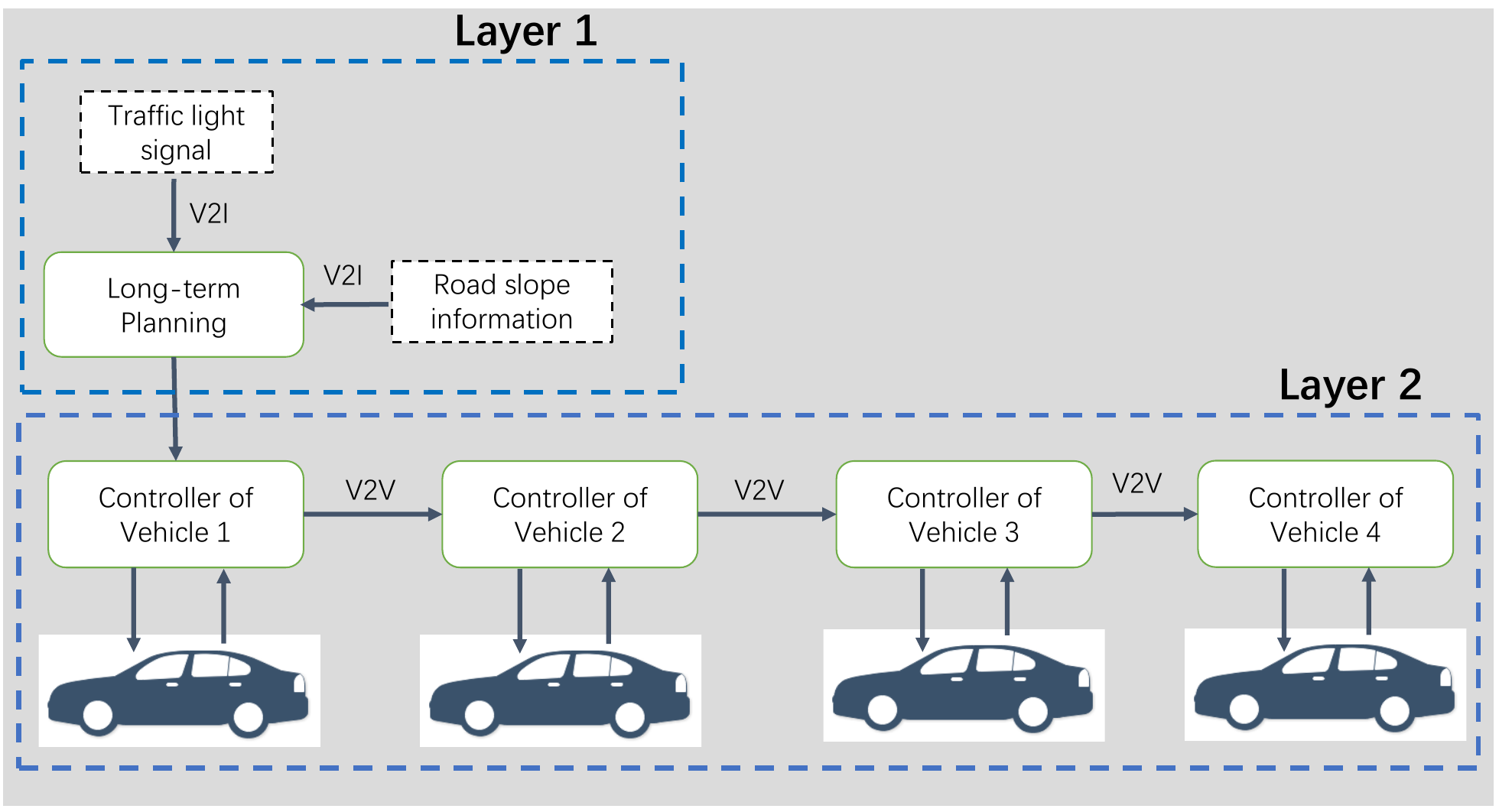}\\
  \caption{Two-layer control architecture.}
\label{fra}
\end{figure}
The framework of the eco-driving algorithm is introduced.
From \eqref{ee4}, we know that the road slope affects the vehicle dynamics. On the other hand, the traffic light information can be used to adjust the vehicle's motion to avoid the red lights. As
Fig.~\ref{fra} shows, in the first layer, the upcoming traffic light signal and the road slope information are used to generate the long-term motion planning for the leader. The long-term planning algorithm runs under an event-triggered mechanism.
In the second layer, the leader designs the real time controller based on the latest long-term motion planning data; and the followers design the real time controller using the motion predictions of the leader and the nearest preceding vehicles. Although the communication topology of layer 2 has been specified in this work, the proposed method is suitable for other appropriate communication  topologies.

The  network unreliability is not considered in the above framework.
This work may lay a foundation for the future research on the vehicle platoon eco-driving
with an unreliable communication network.
\section{Long-term speed profile planning for leader}
In this section, the long-term planning problem is formulated, and the corresponding solution algorithm is developed.


\subsection{Reachability Analysis}
\label{la}
To avoid stopping at red lights can significantly improve fuel efficiency.
When the platoon is close to the traffic light, a new long-term planning for the upcoming traffic light will be defined and triggered. The triggered time of long-term planning for traffic light $j$ is denoted by $t^{[j]}$ satisfying
\begin{align}
\label{qq20}
s_{1}(t^{[j]})\approx p_{j}-\frac{3}{5}c_{j}v_{\max},
\end{align}
where $p_{j}$ is the position of the traffic light $j$.
At time $t^{[j]}$, the leader firstly needs to analyze whether the platoon can go through the traffic light $j$ at the green light windows under the speed limits \eqref{b15}. Based on the  reachability analysis,
the corresponding position constraints induced by the traffic light will be developed.
Firstly, we make the following assumption:
\begin{assumption}
All the considered vehicles always travel as a platoon.
The platoon is allowed to go through the traffic light if the green light will last for longer than $\breve{\tau}$ sampling periods from the time the leader arrives at the traffic light.
\hfill$\blacksquare$
\end{assumption}

The parameter $\breve{\tau}$ mentioned in Assumption 1 is estimated to $\breve{\tau}=\frac{2(N-1)d_{s}}{v_{\min}+v_{\max}}$, where $d_{s}$ is the desired  distance of two adjacent vehicles, $N$ is the number of vehicle in the platoon.
Consider \eqref{z17b} and \eqref{z18b} with Assumption 1. The two feasible upcoming green light windows are modified as:
\begin{itemize}
  \item If $x_{j}(k)=1$, then
\begin{align}
\label{z17}
&\quad[k+t_{j}^{r}-\tau_{j}(k), k+c_{j}-\tau_{j}(k)-\breve{\tau}] \nonumber\\
&\textmd{ and }[k+t_{j}^{r}-\tau_{j}(k)+c_{j}, k+2c_{j}-\tau_{j}(k)-\breve{\tau}].
\end{align}
  \item If $x_{j}(k)=0$, then
\begin{align}
\label{z18}
&\quad[k, k+c_{j}-\tau_{j}(k)-\breve{\tau}]\nonumber\\
&\textmd{ and } [k+t_{j}^{r}-\tau_{j}(k)+c_{j}, k+2c_{j}-\tau_{j}(k)-\breve{\tau}].
\end{align}
\end{itemize}

To ensure the traffic mobility, the vehicle speed should satisfy the road speed limits \eqref{b15} most of the time. For the upcoming traffic light, the leader has the following three options with priority from high to low:
\begin{enumerate}
  \item go through the traffic light during the first upcoming green window under the speed constraint \eqref{b15}.
  \item go through the traffic light during the second upcoming green window under the speed constraint \eqref{b15}.
  \item stop at the traffic light.
\end{enumerate}
Note that the feasibility of options 1) and 2) needs to be verified through computation.
To avoid stopping at the red light and ensure traffic mobility, our strategy is stated as follows:
\begin{strategy}
The leader will choose a feasible option with highest priority from the above options 1)-3) based on the computation.
\end{strategy}

\begin{figure}[!t]
  \centering
  \includegraphics[width=0.7\linewidth]{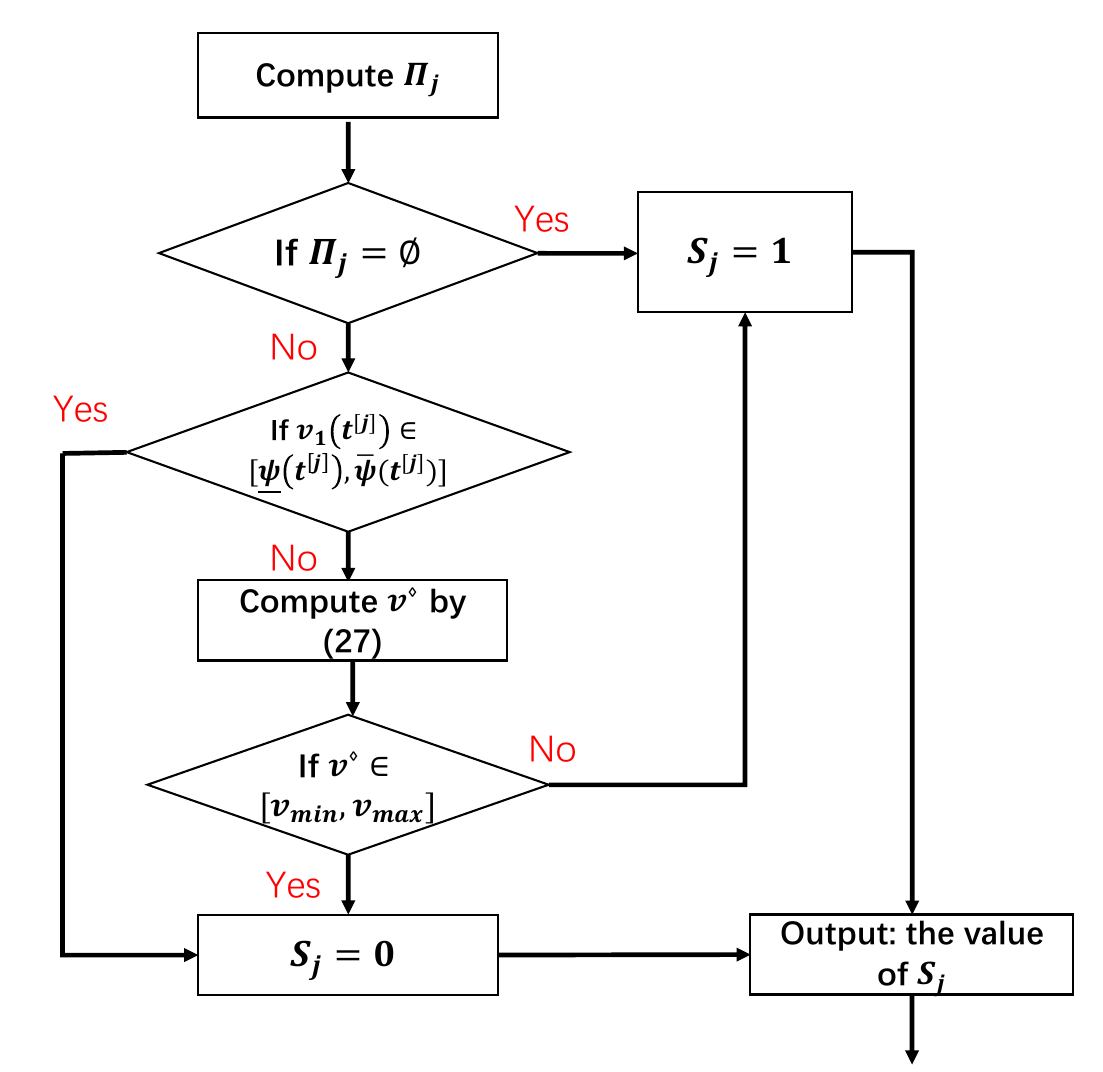}\\
 \caption{The procedure of computing the value of $S_{j}$, where $S_{j}=0$ means that the platoon can go through the traffic light $j$ during the green light windows, and $S_{j}=1$ implies that the platoon has to stop at traffic light $j$.}
\label{flow}
\end{figure}

For simplicity of presentation, a binary variable $S_{j}$ is defined as:
\begin{align}
\label{zz21}
\mathcal{S}_{j}=\left\{
                  \begin{array}{ll}
                    0, & \hbox{if option 1) or 2) is feasible} \\
                    1, & \hbox{otherwise}
                  \end{array}
                \right.
\end{align}
where $\mathcal{S}_{j}=0$ means that the platoon can avoid red light $j$;  $\mathcal{S}_{j}=1$ means that the platoon needs to stop at traffic light $j$.
The value of $S_{j}$ is computed by the procedure summarized in Fig.~\ref{flow} at $k=t^{[j]}$. The details of each part of the procedure in Fig.~\ref{flow} are presented in the remainder of this subsection.

Consider \eqref{z17}, \eqref{z18}. The feasible speed range with respect to the first upcoming green light window is $[q_{1},~q_{2}]$,
where $q_{1}=\frac{\ell_{j}}{c_{j}-\tau_{j}(k)-\breve{\tau}}$, $q_{2}=\left\{
  \begin{array}{ll}
   \breve{q}_{2} &  x_{j}(t^{[j]})=1, \\
   +\infty  & x_{j}(t^{[j]})=0,
  \end{array}
\right.$, $\breve{q}_{2}=\frac{\ell_{j}}{t^{r}_{j}-\tau_{j}(k)}$, $\ell_{j}=|p_{j}-s_{1}(t^{[j]})|$. Note that if the leader's
average speed from $s_{1}(t^{[j]})$ to $p_{j}$ is smaller than $q_{1}$ or bigger than $q_{2}$, then the platoon is not allowed to go through the traffic light $j$ at the first upcoming green light window.
Similarly, the feasible speed range with respect to the second upcoming green window is $[q_{3},~q_{4}]$,
where $q_{3}=\frac{\ell_{j}}{2c_{j}-\tau_{j}(k)-\breve{\tau}}$ and $q_{4}=\frac{\ell_{j}}{t^{r}_{j}-\tau_{j}(k)+c_{j}}$.
Note that the platoon should subject to the road speed limits $[v_{\min},~v_{\max}]$ most of the time. Recalling the feasible speed ranges for the green light  windows, we have that the platoon has to stop at traffic light $j$ if $[v_{\min},~v_{\max}]\cap[q_{1},~\tilde{q}_{2}]=\emptyset$ and $[v_{\min},~v_{\max}]\cap[q_{3},~q_{4}]=\emptyset$. That is
\begin{align}
\label{qq21}
 S_{j}=1 \textmd{ if } \Pi_{j}=\emptyset,
\end{align}
where
$
\Pi_{j}=\Big([q_{1},~q_{2}]\cup[q_{3},~q_{4}]\Big)\cap [v_{\min},~v_{\max}]
$.
For the case $\Pi_{j}\neq\emptyset$,  under Strategy 1, the feasible speed range for going through the traffic light $j$ is  $[\underline{\psi}^{[j]},~\bar{\psi}^{[j]}]$, where
\begin{align}
&\underline{\psi}^{[j]}\nonumber\\
&=\left\{
                            \begin{array}{ll}
                            \max(q_{1},v_{\min}) & \textmd{if } [v_{\min},v_{\max}]\cap[q_{1},\tilde{q}_{2}]\neq\emptyset,   \\
                            \max(q_{3},v_{\min}) & \textmd{otherwise},
                            \end{array}
                          \right.\\
&\bar{\psi}^{[j]}\nonumber\\
&=\left\{
                           \begin{array}{ll}
                            \min(\tilde{q}_{2},v_{\max}) & \textmd{if } [v_{\min},v_{\max}]\cap[q_{1},\tilde{q}_{2}]\neq\emptyset,\\
                            \min(q_{4},v_{\max}) & \textmd{otherwise}. \\
                           \end{array}
                         \right.
\end{align}
For $\Pi_{j}\neq\emptyset$, if $v_{1}(t^{[j]})\in[\underline{\psi}^{[j]},~\bar{\psi}^{[j]}]$, obviously, the platoon can avoid red light:
\begin{align}
\label{qq25}
 S_{j}=0 \textmd{ if } \Pi_{j}\neq\emptyset, \textmd{ and } v_{1}(t^{[j]})\in[\underline{\psi}^{[j]},~\bar{\psi}^{[j]}].
\end{align}
The time window corresponding to the speed range $[\underline{\psi}^{[j]},~\bar{\psi}^{[j]}]$ is
\begin{align}
\label{gw}
\Big[t^{[j]}+\frac{\ell_{j}}{\bar{\psi}^{[j]}},~t^{[j]}+\frac{\ell_{j}}{\underline{\psi}^{[j]}}\Big].
\end{align}
For the case $v_{1}(t^{[j]})\notin[\underline{\psi}^{[j]},~\bar{\psi}^{[j]}]$, we need to analyze whether
the platoon can arrive at the traffic light $j$ during the time range \eqref{gw}.
Considering the smoothness of the driving and the simplicity of analysis, we use the following case for the reachability analysis:
the leader drives with a constant acceleration
\begin{align}
\label{zq29}
 \breve{a}=\left\{
            \begin{array}{ll}
              \frac{3\bar{F}_{1,T}}{5M_{1}}, & \hbox{if $v_{1}(t^{[j]})<\underline{\psi}^{[j]}$}, \\
              -\frac{3\bar{F}_{1,T}}{5M_{1}}, & \hbox{if $v_{1}(t^{[j]})>\bar{\psi}^{[j]}$},
            \end{array}
          \right.
\end{align}
to adjust the speed from $v_{1}(t^{[j]})$ to $v^{\diamond}$, and then drives with a constant speed $v^{\diamond}$ until arriving at traffic light $j$.
Here, if $v_{1}(t^{[j]})<\underline{\psi}^{[j]}$, the leader will accelerate to $v^{\diamond}>v_{1}(t^{[j]})$, while, if
$v_{1}(t^{[j]})>\bar{\psi}^{[j]}$, the leader will decelerate to $v^{\diamond}<v_{1}(t^{[j]})$.
To ensure that the leader arrives at the traffic light during the time window given in \eqref{gw} (i.e., no later than $t^{[j]}+\frac{\ell_{j}}{\underline{\psi}^{[j]}}$ if $v_{1}(t^{[j]})<\underline{\psi}^{[j]}$, and no earlier than $t^{[j]}+\frac{\ell_{j}}{\bar{\psi}(t^{[j]})}$ if $v_{1}(t^{[j]})>\bar{\psi}(t^{[j]})$). Then,
$v^{\diamond}$ should satisfy
\begin{align}
\label{qq27}
  &\frac{(v^{\diamond}-v_{1}(t^{[j]}))(v^{\diamond}+v_{1}(t^{[j]}))}{2\breve{a}}\nonumber\\
&\quad\quad\quad+v^{\diamond}(\breve{t}-\frac{(v^{\diamond}-v_{1}(t^{[j]})}{\breve{a}})=\ell_{j},
\end{align}
where
$\breve{t}=\left\{
           \begin{array}{ll}
            \frac{\ell_{j}}{\underline{\psi}^{[j]}}, & \hbox{if $v_{1}(t^{[j]})<\underline{\psi}^{[j]}$}, \\
             \frac{\ell_{j}}{\bar{\psi}^{[j]}}, & \hbox{if $v_{1}(t^{[j]})>\bar{\psi}^{[j]}$},
           \end{array}
         \right.$.
In the meantime, $v^{\diamond}$ should satisfy the road speed limits $v^{\diamond}\in[v_{\min},v_{\max}]$. Thus, we have the following results:
\begin{align}
\label{qq28}
S_{j}&=0, \textmd{if } \Pi\neq\emptyset,  \textmd{ and } v_{1}(t^{[j]})\notin[\underline{\psi}^{[j]},~\bar{\psi}^{[j]}],  \nonumber\\
&\quad\quad\quad\textmd{ and }v^{\diamond}\in[v_{\min},v_{\max}].\\
\label{qq29}
S_{j}&=1, \textmd{if } \Pi\neq\emptyset, \textmd{ and } v_{1}(t^{[j]})\notin[\underline{\psi}^{[j]},~\bar{\psi}^{[j]}],\nonumber\\
&\quad\quad\quad\textmd{ and } v^{\diamond}\notin[v_{\min},v_{\max}].
\end{align}
Therefore, the value of $S_{j}$ is determined by \eqref{qq21}, \eqref{qq25}, \eqref{qq28}, \eqref{qq29}, as
Fig.~\ref{flow} presents.

Now, the position constraints induced by traffic light $j$ for the leader are defined:

\begin{align}
\label{15-33}
\left\{
  \begin{array}{ll}
    s_{1}(\underline{t}^{[j]})\leq p_{j},\quad
s_{1}(\bar{t}^{[j]})\geq p_{j}, & \hbox{if $S_{j}=0$} \\
  s_{1}(\underline{k}^{[j]})\leq p_{j},\quad s_{1}(\bar{k}^{[j]})\geq p_{j}  , & \hbox{if $S_{j}=1$}
  \end{array}
\right.,
\end{align}
where $\underline{t}^{[j]}=t^{[j]}+\frac{\ell_{j}}{\bar{\psi}^{[j]}}$; $\bar{t}^{[j]}=t^{[j]}+\frac{\ell_{j}}{\underline{\psi}^{[j]}}$;
$\underline{k}^{[j]}=t^{[j]}+2c_{j}+t_{j}^{r}-\tau_{j}(t^{[j]})$, $\bar{k}^{[j]}=t^{[j]}+3c_{j}-\tau_{j}(t^{[j]})-\breve{\tau}$. The existence of the solution space of the constraint \eqref{15-33} follows from the definitions of $\underline{t}^{[j]}$, $\bar{t}^{[j]}$, $\underline{k}^{[j]}$, $\bar{k}^{[j]}$, directly.
The constraint \eqref{15-33} means that 1) if $S_{j}=0$, the leader should go through the traffic light $j$ during the green light window $[\underline{t}^{[j]},~\bar{t}^{[j]}]$; 2) if $S_{j}=1$, the leader should stop at the traffic light until the third green light window $[\underline{k}^{[j]},~\bar{k}^{[j]}]$ comes.

\subsection{Long-term planning problem}
In this subsection, the long-term planning model is developed, and its operation mechanism called event-triggered mechanism is defined.
 Firstly, we define the cost function of the long-term planning problem as follows:
\begin{align}
\label{19-33}
J&=\sum_{k=k_{0}}^{k_{0}+K-1}\Big(\gamma_{P}P_{1}(k)\Delta t+\gamma_{B}(F_{1,B})^{2}\Big) \nonumber\\
&+\sum_{k=k_{0}}^{k_{0}+K}\Big(\gamma_{v}(v_{1}(k)-v_{ref}(k))^{2}+\gamma_{s}(s_{1}(k)-s^{\star}(k))^{2}\Big),
\end{align}
where $v_{ref}$ is the desired speed; and $s^{\star}(k)=s_{1}(k_{0})+(k-k_{0})v_{\max}$. For the cost function $J$, the first term is with respect to (w.r.t.) fuel consumption; the second term is used to reduce the unnecessary braking; the third term is used to maintain the speed within the neighborhood of the desired values; the last term is used to drive the vehicle forward.
Then, the long-term planning model is formulated as follows:
\begin{problem}
 \begin{align*}
\min_{F_{1,T},~F_{1,B}}& J \\
\textmd{s.t.}\quad &\eqref{eq1}\textmd{--}\eqref{eq8},~\eqref{zeq10}\textmd{--}\eqref{zz12};\\
& \eqref{15-33} \textmd { if }  k_{0}=t^{[j]},~j\in \mathds{N}_{+},
\end{align*}
\end{problem}
where $\mathds{N}_{+}$ is the set of positive integer. The time horizon $K$ should be reasonably chosen for each long-term planning. For a long-term planning model with $k_{0}=t^{[j]}$, we need to choose $K$ satisfying $k_{0}+K\geq\bar{t}^{[j]}$ if $S_{j}=0$, and $k_{0}+K\geq \bar{k}^{[j]}$ if $S_{j}=1$.

Now, the operation mechanism (called event-triggered mechanism) for operating the long-term planning model is proposed.
A new long-term planning will be triggered when the following condition holds:
\begin{align}
t\geq k_{0}+K-\varrho   \textmd{ or }  t =t^{[j]},~j\in \mathds{N}_{+},
\end{align}
where $t$ represents the current time; $k_{0}$ and $K$ are the initial time and the horizon of the latest long-term panning, respectively; $\varrho$ is a constant that $\varrho\ll K$. For the triggered condition: 1) $t\geq k_{0}+K-\varrho $ means that the long-term planning data is about to run out, and thus a new long-term planning needs to be triggered; 2)  $t =t^{[j]}$ implies that the platoon is close to the traffic light, and it needs a new long-term planning for avoiding the upcoming red light.

\subsection{Solution Algorithm}
Define $u=[F_{1,T},~F_{1,B}]$, $x=[v_{1},~s_{1}]$, $z=[u_{1}(k),x_{1}(k+1),\ldots,u_{1}(k+K-1),x_{1}(k+K)]\tr$, and
\begin{align*}
\mu_{1}=\left\{
          \begin{array}{ll}
            \underline{t}^{[j]}-t^{[j]}, & \hbox{if $k_{0}=t^{[j]}$, $S_{j}=0$;} \\
         \underline{k}^{[j]}-t^{[j]}, & \hbox{if $k_{0}=t^{[j]}$, $S_{j}=1$;} \\
 K+1, & \hbox{otherwise;} \\
          \end{array}
        \right.\\
\mu_{2}=\left\{
          \begin{array}{ll}
            \bar{t}^{[j]}-t^{[j]}, & \hbox{if $k_{0}=t^{[j]}$, $S_{j}=0$;} \\
         \bar{k}^{[j]}-t^{[j]}, & \hbox{if $k_{0}=t^{[j]}$, $S_{j}=1$;} \\
 K+1, & \hbox{otherwise.} \\
          \end{array}
        \right.
\end{align*}
Let $I_{n}$ denote the $n\times n$ unit matrix, and $0_{n\times m}$ denote the $n\times m$ matrix whose all elements are zero.
Normalize the time coordinates to $k_{0}=0$.
Then, the optimization problem as Problem 1 can be written as the following compact form:
 \begin{align}
\label{p34}
 \left\{
   \begin{array}{ll}
\min\limits_{z}\quad &J=\frac{1}{2}z\tr Hz+b\tr z+c \\
\textmd{s.t.}\quad &Gz\leq h,~~Ez=d.
   \end{array}
 \right.
\end{align}
where, $H=2\times\textmd{diag}\{H_{1},H_{2},\ldots,H_{K+1}\}$, $b=[b_{1},~\cdots,b_{K+1}]\tr$, $G=\textmd{diag}\{\bar{G},G_{1},\bar{G},G_{2},\cdots,\bar{G},G_{T}\}$, $h=[\bar{h},h_{1},\bar{h},h_{2},\ldots,\bar{h},h_{K}]\tr$,  $E=\Big[E_{1}\tr,\ldots,E_{K}\tr\Big]\tr$, $d=[(A_{0}x\tr(0))\tr,\quad 0_{1\times 2(K-1)}]\tr$, and
\begin{align*}
H_{1}&=\begin{bmatrix}
      \varsigma_{11} & \\
         & \gamma_{B} \\
      \end{bmatrix},~H_{K+1}=\begin{bmatrix}
      \gamma_{v} & \\
         & \gamma_{s}\\
      \end{bmatrix},\\
H_{\mu}&=\begin{bmatrix}
         \gamma_{v}+\varsigma_{22} & 0 & \varsigma_{21} & 0 \\
         0 &\gamma_{s} & 0 & 0 \\
         \varsigma_{12} & 0 & \varsigma_{11} & 0 \\
         0 & 0 &  0&  \gamma_{B} \\
       \end{bmatrix},~\mu\in\{2,\ldots,K\},\\
\varsigma_{ij}&=o_{1,ij}\times \gamma_{P}\Delta t, ~\bar{\varsigma}_{i}=\bar{o}_{1,i}\times \gamma_{P}\Delta t,~i,j\in\{1,2\},\\
b_{1}&=\Big[2\varsigma_{21}v_{1}(0)+\bar{\varsigma}_{1},~0\Big],\\
b_{\mu}&=\Big[-2\gamma_{v}v_{ref}(\mu-1)+\bar{\varsigma}_{2},~-2\gamma_{s}s^{\star}(\mu-1),\\
&\quad\quad~\bar{\varsigma}_{1},~0 \Big],~~\mu\in\{2,\ldots,K\},\\
b_{T+1}&=\Big[-2\gamma_{v}v_{ref}(T)+\bar{\varsigma}_{2},~-2\gamma_{s}s^{\star}(T)\Big],\\
\bar{G}&=\begin{bmatrix}
          1 & 0\\
          -1 & 0 \\
          0 & 1 \\
          0 & -1 \\
        \end{bmatrix},~~G_{\mu}=\begin{bmatrix}
                                1 & 0 \\
                                -1 & 0 \\
                                0& 1 \\
                                0 & 0 \\
                              \end{bmatrix},~\mu\notin\{\mu_{1},\mu_{2}\},\\
G_{\mu_{1}}&=\begin{bmatrix}
                                1 & 0 \\
                                -1 & 0 \\
                                0& 1 \\
                                0 & 0 \\
                              \end{bmatrix},~~
G_{\mu_{2}}=\begin{bmatrix}
                                1 & 0 \\
                                -1 & 0 \\
                                0& -1 \\
                                0 & 0 \\
                              \end{bmatrix},\\
\bar{h}&=[\bar{F}_{1,T},~0,~\bar{F}_{1,B},~0], \\
h_{\mu}&=[v_{\max},~0,~s_{1}(0)+v_{\max}\mu,~0],~ \mu\notin\{\mu_{1},\mu_{2}\},\\
h_{\mu_{1}}&=[v_{\max},~0,~p_{j},~0],~
h_{\mu_{2}}=[v_{\max},~0,~-p_{j},~0],\\
E_{1}&=[-B~I_{2}~0_{2\times 4(K-1)}],\\
E_{\mu}&=[0_{2\times (4\mu-6)}, -A_{\mu-1},~-B,~I_{2}~0_{2\times 4(K-\mu)}],\\
&\quad\quad~\mu\in\{2,\ldots,K\},\\
B&=\begin{bmatrix}
    \frac{\Delta t}{M_{1}} &  -\frac{\Delta t}{M_{1}} \\
    \frac{(\Delta t)^{2}}{2M_{1}} &  -\frac{(\Delta t)^{2}}{2M_{1}}  \\
  \end{bmatrix},\\
A_{\mu}&=\begin{bmatrix}
         1-\frac{\Delta t}{M}\xi_{1,d}v_{1}(\mu)   &-\frac{\Delta t}{Ms_{1}(\mu)}\bar{F}_{1,E}(\mu)  \\
         \Delta t-\frac{(\Delta t)^{2}}{2M}\xi_{1,d}v_{1}(\mu)   & 1-\frac{(\Delta t)^{2}}{2Ms_{1}(\mu)}\bar{F}_{1,E}(\mu) \\
         \end{bmatrix},\\
\bar{F}_{1,E}(\mu)&=g\sin(\theta(s_{1}(\mu)))+gc_{1}^{r}\cos(\theta(s_{1}(\mu))).~~~~~~~~~~
\textmd{\hfill$\blacksquare$}
\end{align*}
A constraint without changing the solution space of Problem 1 (i.e., $s_{1}(k)\leq s_{1}(0)+v_{\max}k$) is added into the model \eqref{p34} such that the matrix $G$ is column full rank.
The matrix $E$ and the vector $d$ are state-dependent parameters. We aim to solve problem \eqref{p34} by the numerical algorithm. In the iteration process, for
a given $z$, the matrix $E$ and the vector $d$ are considered to be constant.
To solve problem \eqref{p34}, the following augmented Lagrangian function is defined:
\begin{align}
\mathcal{L}(z,\lambda,\eta)&=z\tr H z+b\tr z+c+\frac{\xi}{2}(Ez-d)\tr(Ez-d)\nonumber\\
&\quad+\lambda\tr(Ez-d)+\eta\tr(Gz-h+\pi),\nonumber\\
&=z\tr \bar{H} z+\bar{b}\tr z+\bar{c}+\lambda\tr(Ez-d)\nonumber\\
&\quad+\eta\tr(Gz-h+\pi),
\end{align}
where $\xi$ is a big positive number;
$\bar{H}=H+\frac{\xi}{2}E\tr E$, $\bar{b}=b-\xi E\tr d$, $\bar{c}=c+\frac{\xi}{2}d\tr d$.
The KKT condition of problem \eqref{p34} is given by
\begin{align}
\label{28-42}
\mathcal{F}(z,\lambda,\eta,\pi)=\begin{bmatrix}
                                  \bar{H}z+\bar{b}+E\tr \lambda+G\tr \eta \\
                                  Ez-d \\
                                  Gz-h+\pi \\
                               \Pi \eta ~(\textmd{or }\Psi\pi)  \\
                                \end{bmatrix}=0_{(22K)\times 1},
\end{align}
where $\Pi=diag(\pi)$, $\Psi=diag(\eta)$. The Jacobian matrix is given by
\begin{align*}
\mathcal{J}_{ac}=\begin{bmatrix}
                                  \frac{\partial\mathcal{F}}{\partial z\tr} &
                                  \frac{\partial\mathcal{F}}{\partial \lambda\tr} &
                                   \frac{\partial\mathcal{F}}{\partial \eta\tr} &
                                   \frac{\partial\mathcal{F}}{\partial \pi\tr} \\
                                \end{bmatrix}
=\begin{bmatrix}
                                  \bar{H}  & E\tr & G\tr & 0 \\
                                  E & 0 &0  & 0 \\
                                  G & 0 & 0 & I \\
                                  0 & 0 & \Pi & \Psi \\
                                \end{bmatrix}.
\end{align*}
Applying the Newton's method to solve the KKT condition \eqref{28-42}, the search direction ($ \Delta z, \Delta\lambda,\Delta\eta, \Delta \pi $) is given by solving the following linear equations:
\begin{align}
\label{07-38}
\mathcal{J}_{ac}\begin{bmatrix}
                                               \Delta z \\
                                              \Delta\lambda \\
                                               \Delta\eta \\
                                               \Delta \pi \\
                                             \end{bmatrix}=\begin{bmatrix}
                                                                                             r_{z} \\
                                                                                             r_{\lambda}\\
                                                                                             r_{\eta} \\
                                                                                             r_{\pi} \\
                                                                                           \end{bmatrix}
=-\mathcal{F}(z,\lambda,\eta,\pi).
\end{align}
It follows from $\Pi\Delta \eta+\Psi\Delta \pi=r_{\pi}$ that
$\Delta \pi=\Psi^{-1}(r_{\pi}-\Pi\Delta\eta)$. Thus, $G \Delta z+\Delta\pi=r_{\eta}\Longleftrightarrow G \Delta z-\Psi^{-1}\Pi\Delta\eta=r_{\eta}-\Psi^{-1}r_{\pi}$. Then, \eqref{07-38} is equivalent to:
\begin{align}
\label{07-39}
\left\{
  \begin{array}{ll}
\begin{bmatrix}
                                  \bar{H}  & E\tr& G\tr \\
                                  E & 0 &0   \\
                                  G & 0 &-\Psi^{-1}\Pi  \\
                                \end{bmatrix}\begin{bmatrix}
                                               \Delta z \\
                                              \Delta\lambda \\
                                               \Delta\eta \\
                                             \end{bmatrix}=\begin{bmatrix}
                                                                                             r_{z} \\
                                                                                             r_{\lambda}\\
                                                                                             r_{\eta}-\Psi^{-1} r_{\pi} \\
                                                                                           \end{bmatrix} \\
\Delta \pi=\Psi^{-1}(r_{\pi}-\Pi\Delta\eta)
  \end{array}
\right..
\end{align}
From \eqref{07-39}, one has $\Delta \eta=\Pi^{-1}\Psi(G \Delta z-r_{\eta}+\Psi^{-1}r_{\pi})$, which
implies that $\bar{H} \Delta z+E\tr \Delta \lambda +G\tr \Delta \eta=r_{z}\Longleftrightarrow (\bar{H} +G\tr \Pi^{-1}\Psi G)\Delta z+E\tr \Delta\lambda=r_{z}+G\tr\Pi^{-1}\Psi r_{\eta}-G\tr\Pi^{-1}r_{\pi}\triangleq \bar{r}$.
As a result, \eqref{07-39} is reduced to
\begin{align}
\label{28-47}
\left\{
  \begin{array}{ll}
   \begin{bmatrix}
  \bar{H} +G\tr\Pi^{-1}\Psi G & E\tr \\
 E  & 0 \\
\end{bmatrix}\begin{bmatrix}
               \Delta z \\
               \Delta \lambda \\
             \end{bmatrix}=\begin{bmatrix}
               \bar{r}\\
                r_{\lambda} \\
              \end{bmatrix} \\
   \Delta \eta=\Pi^{-1}\Psi(G \Delta z-r_{\eta}+\Psi^{-1}r_{\pi})\\
  \Delta \pi=\Psi^{-1}(r_{\pi}-\Pi\Delta\eta)
  \end{array}
\right..
\end{align}
Note that $\bar{H}+G\tr \Pi^{-1}\Psi G\triangleq \breve{H}\succ 0$.
Due to that $E$ is row full rank, $E\breve{H}^{-1} E\tr\succ 0$. Equation \eqref{28-47} is simplified to
\begin{align}
\label{28-41}
\left\{
  \begin{array}{ll}
  \Delta \lambda=(E \breve{H}^{-1}E\tr)^{-1}(E \breve{H}^{-1}\bar{r}-r_{\lambda}) \\
\Delta z=\breve{H}^{-1}(\bar{r}-E\tr\Delta \lambda)\\
   \Delta \eta=\Pi^{-1}\Psi(G \Delta z-r_{\eta}+\Psi^{-1}r_{\pi})\\
  \Delta \pi=\Psi^{-1}(r_{\pi}-\Pi\Delta\eta)
  \end{array}
\right..
\end{align}
Now, the Newton's method-based algorithm is summarized (in Algorithm 1) for solving the optimization problem with the form of Problem 1.
\begin{algorithm}
\caption{Calculation of the optimal long-term planning motion}
\begin{algorithmic}[1]
\label{al2}
 \STATE Initial $z\geq0$, $\lambda>0$, $\eta>0$, $\pi>0$.
\STATE Update matrix $E$ and vector $d$ using the current value of $z$.
  \STATE Compute  $[r_{z}~r_{\lambda}~r_{\eta}~r_{\pi}]\tr=-\mathcal{F}(z,\lambda,\eta,\pi)$ based on \eqref{28-42}.
  \IF{ $||\mathcal{F}(z,\lambda,\eta,\pi)||<\epsilon$ ($\varepsilon$ is a small positive number)}
  \STATE The algorithm is stopped.
  \ELSE
  \STATE  Compute ($ \Delta z, \Delta\lambda,\Delta\eta, \Delta \pi $) using \eqref{28-41}.
  \STATE Choose the step length  $\delta=\max\big\{x\in(0,~\varpi]:[\eta~\pi]+x[\Delta\eta~\Delta\pi]>0_{1\times 2}\big\}$, where $\varpi$ is a number smaller than $1$ and close to $1$.
  \STATE Update $[z~\lambda~\eta~\pi]=[z~\lambda~\eta~\pi]+\delta[\Delta z~\Delta\lambda~\Delta\eta~\Delta\pi]$.
  \STATE Return to Step 2.
  \ENDIF
  \STATE \textbf{Output}: The value of $z$.
\end{algorithmic}
\end{algorithm}

\section{Short-Term Adaptation}
\subsection{The leader (vehicle $1$)}
\label{ss1}
After obtaining the long-term planning data $v^{l}_{1}(k),s^{l}_{1}(k)$ ( for $k=k_0,\ldots,k_0+K$) by solving Problem 1, the leader tries to travel following the planning motion profile in real time. The real time controller of the leader is given by solving the following problem:
\begin{problem}
\begin{align*}
\min_{F_{1,T},F_{1,B}} &J_{1}^{r}\nonumber\\
\textmd{s.t. }~
& \eqref{eq1}\textmd{--}\eqref{eq8},~\eqref{zeq10}\textmd{--}\eqref{zz12},
\end{align*}
\end{problem}
where
\begin{align*}
J_{1}^{r}&=\sum_{k=\bar{k}_{0}}^{\bar{k}_{0}+\bar{K}_{1}}\omega_{1,v}(v_{1}(k)-v_{1}^{l}(k))^{2}+\omega_{1,s}(s_{1}(k)-s_{1}^{l}(k))^{2}\nonumber\\
&\quad+\sum_{k=\bar{k}_{0}}^{\bar{k}_{0}+\bar{K}_{1}-1}\omega_{1,p}P_{1}(k)\Delta t+\omega_{1,B}(F_{1,B})^{2},
\end{align*}
here $\omega_{1,v}$, $\omega_{1,s}$, $\omega_{1,p}$ and $\omega_{1,B}$ are weight parameters. The first and second terms of $J_{1}^{r}$ try to take the vehicle to the planning motion profile $(v_{1}^{l}(k), s_{1}^{l}(k))$;  the third term is w.r.t. fuel consumption; the last term is used to reduce the unnecessary braking. Problem 2 can be transformed into the form of \eqref{p34} which can be solved by Algorithm 1.
The real time controller of the leader is summarized in Algorithm 2.
\begin{algorithm}
\caption{Real time controller of the leader}
\begin{algorithmic}[1]
\STATE Let $\bar{k}_{0}=t$, where $t$ is the current time. Measure the current state $v_{1}(t)$, $s_{1}(t)$.
\STATE Use the latest long-term planing data $(v^{l}_{1}(k),s^{l}_{1}(k))$ ( for $k=k_0,\ldots,k_0+K$) to define Problem 2.
\STATE Solve Problem 2 to obtain $F_{1,T}(\bar{k}_{0})$, $F_{1,B}(\bar{k}_{0})$, and $a_{1}(k)$ for $k\in\{\bar{k}_{0}+1,\ldots,\bar{k}_{0}+\bar{K}_{1}\}$.
\STATE Apply $F_{1,T}(\bar{k}_{0})$, $F_{1,B}(\bar{k}_{0})$ to control the vehicle.
\STATE Compute the  predicted values of $v_{1}(k)$, $s_{1}(k)$ recursively based on the vehicle dynamic and $a_{1}(k)$ for $k\in\{\bar{k}_{0},\ldots,\bar{k}_{0}+\bar{K}_{1}\}$, where $a_{1}(k)$ is computed in Step 3.
\STATE Transmit the predicted values of $v_{1}(k)$, $s_{1}(k)$ denoted by $\hat{v}_{1}(k)$, $\hat{s}_{1}(k)$, for $l\in\{\bar{k}_{0},\ldots,\bar{k}_{0}+\bar{K}_{1}\}$ to the followers.
\STATE When the current time become $t+1$, let $t=t+1$, return to Step 1.
\end{algorithmic}
\end{algorithm}

\subsection{The follower (vehicle $i$, $i>1$)}
The follower tries to track the leader and its nearest preceding vehicle such that the distances between the vehicles are  maintained near the desired values. In addition, to avoid collision, the  following condition should hold:
\begin{align}
\label{07-41}
 s_{i}(k)\leq \hat{s}_{i-1}(k)-\Delta s,
\end{align}
where $\Delta s$ is the safe distance designed by the user.
The real time controller of the follower is given by solving the following problem:
\begin{problem}
 \begin{align*}
\min_{F_{i,T}(k),F_{i,B}(k)}~&J_{i}^{r}\\
\textmd{s.t. }~
& \eqref{eq1}\textmd{--}\eqref{eq8},~\eqref{zeq10}\textmd{--}\eqref{zz12},~\eqref{07-41}
\end{align*}
where
\begin{align*}
J_{i}^{r}&=\sum_{k=\bar{k}_{0}}^{\bar{k}_{0}+\bar{K}_{i}}\Big(\omega_{i,v}(v_{i}(k)-\hat{v}_{i-1}(k))^{2}
+\tilde{\omega}_{i,v}(v_{i}(k)-\hat{v}_{1}(k))^{2}\nonumber\\
&\quad+\omega_{i,s}(s_{i}(k)-\hat{s}_{i-1}(k)-d_{s})^{2}\\
&\quad+\tilde{\omega}_{i,s}(s_{i}(k)-\hat{s}_{1}(k)-(i-1) d_{s})^{2}\Big)\\
&\quad+\sum_{k=\bar{k}_{0}}^{\bar{k}_{0}+\bar{K}_{i}-1}\omega_{i,p}P_{i}(k)\Delta t+\omega_{i,B}(F_{i,B})^{2},
\end{align*}
here $i>1$, $\omega_{i,v}$, $\omega_{i,s}$, $\omega_{i,a}$, $\tilde{\omega}_{i,v}$, $\tilde{\omega}_{i,s}$, $\omega_{i,P}$ and $\omega_{i,B}$ are weight parameters; and
$\tilde{\omega}_{2,v}=\tilde{\omega}_{2,s}=0$; $\hat{v}_{p}(k)$, $\hat{s}_{p}(k)$ ($p\in\{1,~i-1\}$) are the latest received predicted values of the speed and position for vehicle $p$. The first and second terms of $J_{i}^{r}$ are used to track the speeds of the leader and the nearest preceding vehicle; the third and fourth terms are used for maintaining the inter-vehicular distance near the desired values.
\end{problem}

The objective function $J_{i}^{r}$ is defined based on the received information that depends on the communication topology. In this paper, a leader-predecessor-follower communication topology is adopted. However, this work can be applied to other communication topologies just by revising the definition of  $J_{i}^{r}$ accordingly.
Note that the optimization horizon $\bar{K}_{i}$ should be chosen to satisfy $\bar{K}_{i+1}\leq \bar{K}_{i}$, and $\bar{K}_{1}\ll K$ ($K$ appears in Problem 1) such that we have enough data to define Problems 2, 3. The solution of Problem 3 can be obtained by Algorithm 1, which is initialized by  the parameters of Problem 3. The real time controller of the followers is summarized in Algorithm 3.
%
\begin{algorithm}
\caption{Real time controller of the follower}
\begin{algorithmic}[1]
\STATE Let $\bar{k}_{0}=t$, where $t$ is the current time. Measure the current state $v_{i}(t)$, $s_{i}(t)$.
\STATE Use the latest predicted values $\hat{v}_{p}(k)$, $\hat{s}_{p}(k)$, for $p\in\{1,i-1\}$, $k\in\{\bar{k}_{0},\ldots,\bar{k}_{0}+\bar{K}_{i}\}$ to define Problem 3.
\STATE Solve Problem 3 to obtain $F_{i,T}(\bar{k}_{0})$, $F_{i,B}(\bar{k}_{0})$, and $a_{i}(k)$ for $k\in\{\bar{k}_{0}+1,\ldots,\bar{k}_{0}+\bar{K}_{i}\}$.
\STATE Apply $F_{i,T}(\bar{k}_{0})$, $F_{i,B}(\bar{k}_{0})$ to control the vehicle.
\STATE Compute the predicted values of $v_{i}(k)$, $s_{i}(k)$ recursively based on the vehicle dynamic and  $a_{i}(k)$ for $k\in\{\bar{k}_{0},\ldots,\bar{k}_{0}+\bar{K}_{i}\}$.
\STATE Transmit the predicted values of $v_{i}(k)$, $s_{i}(k)$ denoted by $\hat{v}_{i}(k)$, $\hat{s}_{i}(k)$, for $l\in\{\bar{k}_{0},\ldots,\bar{k}_{0}+\bar{K}_{i}\}$ to its nearest followers.
\STATE When the current time become $t+1$, let $t=t+1$, return to Step 1.
\end{algorithmic}
\end{algorithm}
\section{Simulations}
\label{sim}
A platoon composed of $3$ heterogeneous vehicles is considered. The parameters of the vehicle's dynamic are chosen as follows: the vehicle masses  $M_{1}=1420kg$, $M_{2}=1320kg$, $M_{3}=1520kg$; the rolling resistance coefficients $c_{1}^{r}=0.02$, $c_{2}^{r}=0.018$, $c_{3}^{r}=0.022$; the drag coefficients $c^{d}=0.36$; the air density $\rho=1.205kg/m^{3}$; the face area of the vehicle $S_{1,A}=1.7m^{2}$, $S_{2,A}=1.6m^{2}$, $S_{3,A}=1.8m^{2}$; the sampling period $\Delta t=0.5s$.


In \eqref{zeq6}, the air drag coefficients are: $\alpha=0.414$, $\beta=41.29$. Considering  the fuel consumption model \eqref{q9}, the coefficients are chosen to be
$o_{i,11}=1.8085\times 10^{-4}/r_{i}$, $o_{i,12}=o_{i,21}=8.6815\times 10^{-6}/r_{i}$, $o_{i,22}=5.4479\times 10^{-6}/r_{i}^{2}$,
$\bar{o}_{i,1}=0$, $\bar{o}_{i,2}=1.1046\times 10^{-2}/r_{i}$, $\bar{o}_{i,3}=0$.
where $r_{1}=0.30115m$, $r_{2}=0.29915m$, $r_{3}=0.31015m$ are the tire radius of the vehicles. For the coefficients given above, the unit of the fuel consumption rate $P_{i}$ is $10^{-6}$ $kg/s$. The parameters in \eqref{zeq10}--\eqref{b15} are set as $\bar{F}_{i,T}=(6.5\times M_{i})[N]$, $\bar{F}_{i,B}=(4\times M_{i})[N]$,
$v_{\min}=8m/s$, $v_{\max}=16m/s$.
The road profile is described by the following piecewise function:
\begin{align}
h(s)=\left\{
       \begin{array}{ll}
        -\frac{1}{20000}s^{2}+\frac{1}{100}s , & \textmd{ if } 0\leq s\leq 200;\\
        \frac{1}{5000}s^{2}-\frac{1}{10}s+12 , & \textmd{ if } 200< s\leq 300; \\
        -\frac{1}{10000}s^{2}+\frac{2}{25}s-15 , & \textmd{ if } 300< s\leq 500;\\
0.1s+500,              & \textmd{ if } s>500.\\
       \end{array}
     \right.
\end{align}
Then, the road slope is given by $\theta(s)=\frac{\partial h(s)}{\partial s}$:
\begin{align}
 \theta(s)=\left\{
       \begin{array}{ll}
        -\frac{1}{10000}s+\frac{1}{100} , & \textmd{ if } 0\leq s\leq 200;\\
        \frac{1}{2500}s-\frac{1}{10} , & \textmd{ if } 200< s\leq 300; \\
        -\frac{1}{5000}s+\frac{2}{25} , & \textmd{ if } 300< s\leq 500;\\
0.1,            & \textmd{ if } s>500.\\
       \end{array}
     \right.
\end{align}
\subsection{Long-term planning}
\label{ltp}
For the upcoming traffic light $j$, the red and green light durations are $t_{j}^{r}=20s$, $t_{j}^{g}=7s$, respectively. Then, the clock signal period of the traffic light $j$ is $c_{j}=27s$. The estimation of the time for the platoon going through the traffic light is  $\breve{\tau}=\frac{2(N-1)d_{s}}{v_{\min}+v_{\max}}=0.5$s, where the desired distance of two
adjacent vehicles is chosen to $d_{s}=3$ m. The position of the traffic lights are $p_{1}=260$ m, $p_{2}=580$ m, $p_{3}=980$ m.
The initial speed and position of the leader are $v_{1}(0)=9$ m/s, $s_{1}(0)=10$ m. The weight parameters in the cost function \eqref{19-33} are set to
$\gamma_{P}=0.4$, $\gamma_{B}=0.01$, $\gamma_{v}=0.7$, $\gamma_{s}=0.001$. The desired speed is set as:
\begin{itemize}
\item if $k_{0}=t^{[j]}$ and $S_{j}=0$, $v_{ref}(k)=\left\{
                                          \begin{array}{ll}
                                          \min(\breve{v}(k),\bar{\psi}^{[j]}-1) , & \hbox{if $v_{1}(k_{0})<\underline{\psi}^{[j]}$} \\
                                          \max(\breve{v}(k),\bar{\psi}^{[j]}-3)  , & \hbox{if $v_{1}(k_{0})>\bar{\psi}^{[j]}$} \\
                                          \max(\breve{v}(k),\bar{\psi}^{[j]}-1)  , & \hbox{otherwise}
                                          \end{array}
                                        \right.$, where $\breve{v}(k)=v_{1}(k_{0})+(k-k_{0})(\bar{\psi}^{[j]}-v_{1}(k_{0}))/K$.

  \item  if $k_{0}=t^{[j]}$ and $S_{j}=1$, $v_{ref}(k)=v_{1}(k_{0})-(v_{1}(k_{0})/K)(k-k_{0})$.
\item if $k_{0}\neq t^{[j]}$, $j\in \mathds{N}_{+}$, then $v_{ref}(k)=13.3$ m/s.
\end{itemize}
The time horizon of the long-term planning is set to be $K=\textmd{max}.(\mu_{2},64)$ if $k_{0}=t^{[j]}$, otherwise, $K=64$.
The long-term planning problem formulated as Problem 1 is solved by Algorithm 1. The computation time is about $5$ seconds for $K=64$ (running by MATLAB R2020b
on Intel(R) Core(TM) i5-3337U CPU (1.80GHz)). The computation time is reduced to about $0.3$ s if $K=24$. Inspired by this, we can decompose Problem 1 with time horizon $K$ into two subproblems with time horizons $K_{1}\leq 24$ and $K_{2}=K-K_{1}$, respectively. Then, the long-term planning data can be available in time to the leader. In this simulation, for simplicity of the algorithm implement, we assume that the platoon can obtain the long-term planning data in time by solving Problem 1 directly.

\begin{figure}[!h]
  \centering
  \includegraphics[width=0.7\linewidth]{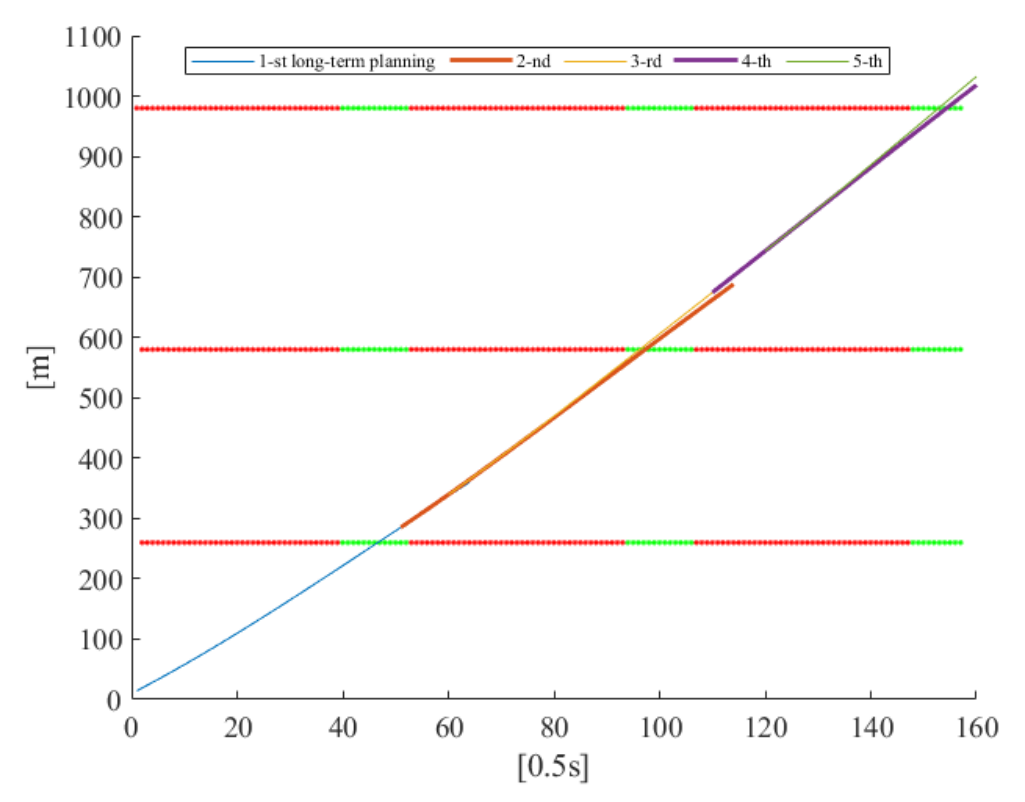}\\
  \caption{The position profile of long-term planning.}
\label{l_p}
\end{figure}

\begin{figure}[!h]
  \centering
  \includegraphics[width=0.7\linewidth]{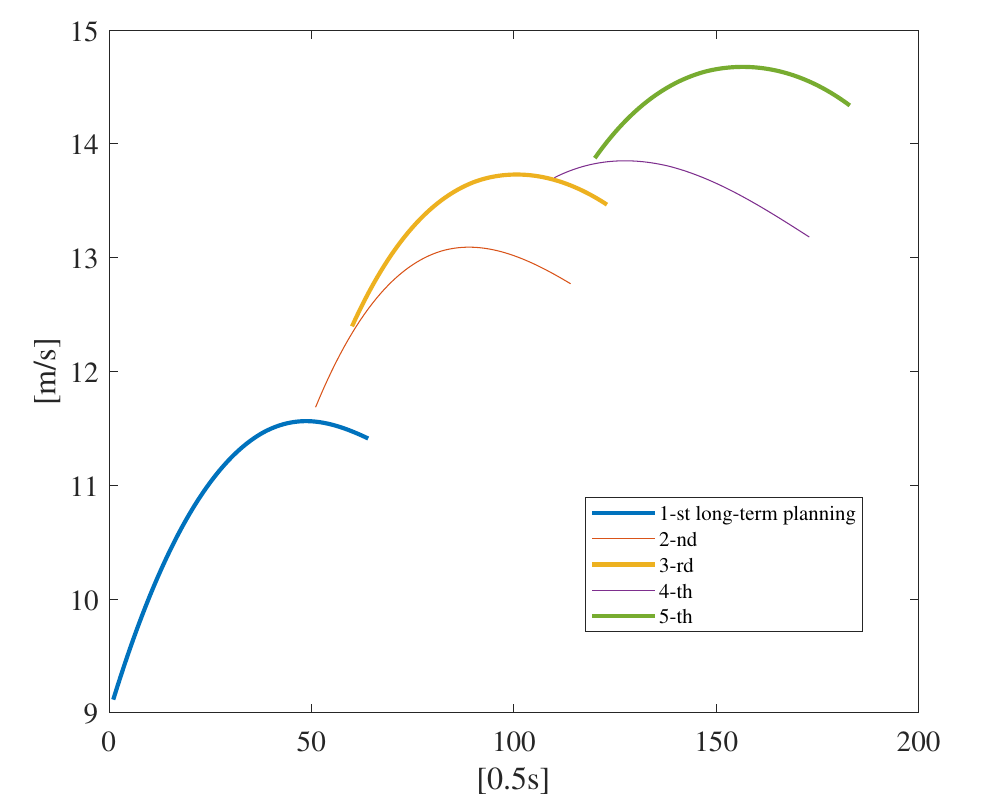}\\
  \caption{The speed profile of long-term planning.}
\label{l_s}
\end{figure}

The position and speed profiles of the long-term planning for the leader are shown in Figs.~\ref{l_p}--\ref{l_s}. Fig.~\ref{l_p} shows that the
leader can avoid red lights if driving exactly as planned. Note
that the feasible green light duration is $t_{j}^{g}-\breve{\tau}=6.5$ s,
and the
red light duration is $t_{j}^{r}=20$ s.
The proposed algorithm can
effectively find the feasible speed/position profiles to avoid
the red lights under the speed limits.

\subsection{Real Time Adaption}
\label{rta}
\begin{figure}[!h]
  \centering
  \includegraphics[width=0.7\linewidth]{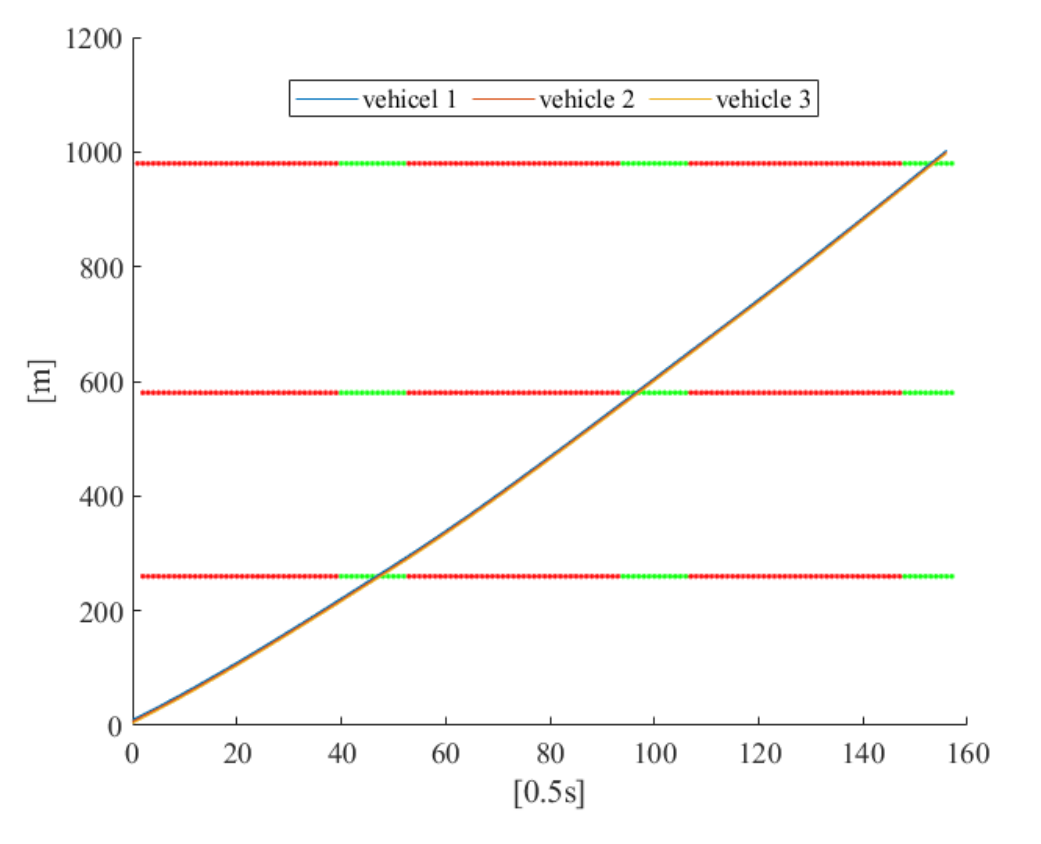}\\
  \caption{The real time position profile.}
\label{r_p}
\end{figure}

\begin{figure}[!h]
  \centering
  \includegraphics[width=0.7\linewidth]{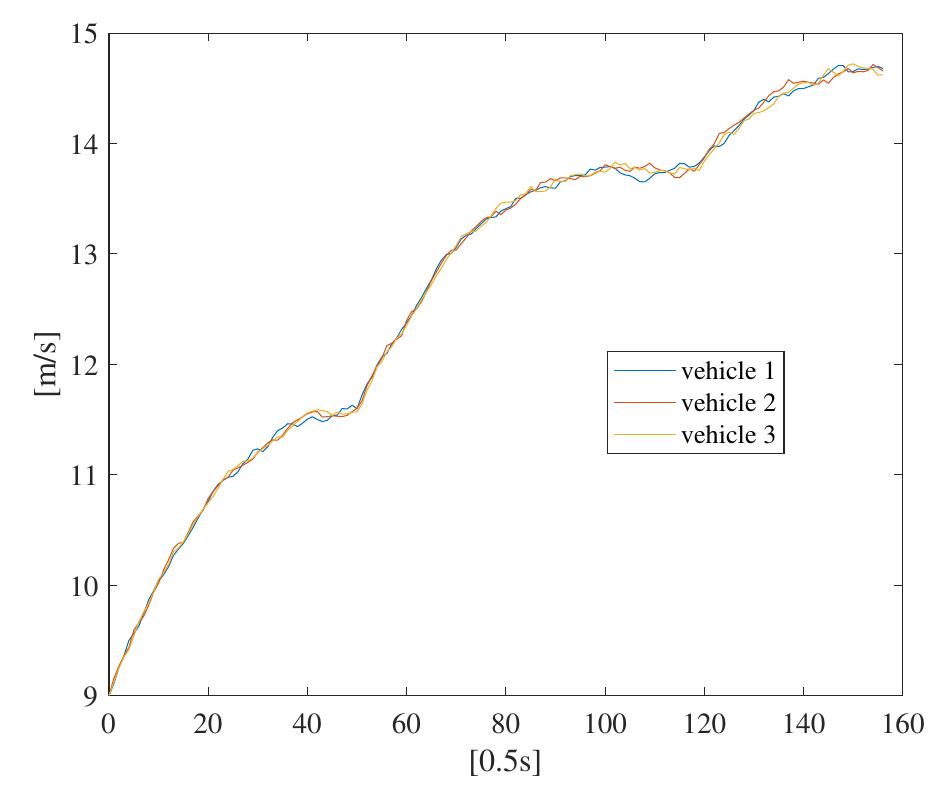}\\
  \caption{The real time speed profile.}
\label{r_s}
\end{figure}

\begin{figure}[!h]
  \centering
  \includegraphics[width=0.7\linewidth]{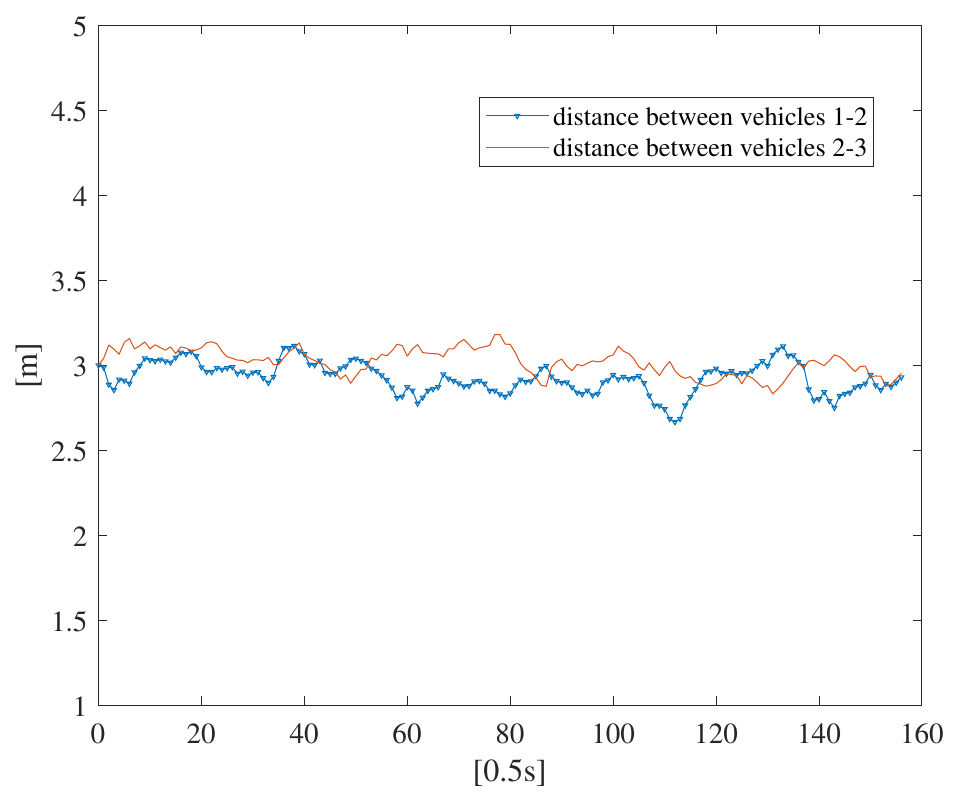}\\
  \caption{The inter-vehicular distances.}
\label{r_e}
\end{figure}

For real time driving, the modeling errors of the vehicle dynamics are considered. In particular, the speed error and the position error are zero-mean Gaussian variables with standard deviation $0.02$ m/s and $0.02$ m, respectively. The real time controller is obtained by Algorithms 2--3, in which
Problems 2--3 should be solved. The parameters of Problems 2--3 are given as follows:
$\bar{K}_{1}=12$, $\bar{K}_{2}=12$, $\bar{K}_{3}=12$, $\omega_{1,v}=2$, $\omega_{1,s}=2$, $\omega_{1,B}=0.01$, $\omega_{1,P}=0.1$;
$\omega_{2,v}=\omega_{3,v}=6$, $\omega_{2,s}=\omega_{3,s}=6$, $\omega_{2,B}=\omega_{3,B}=0.01$, $\omega_{2,P}=\omega_{3,P}=0.06$.
The desired inter-vehicular distance is $d_{s}=3$.
%
The real time position and speed profiles of the platoon are given in Figs.~\ref{r_p}-\ref{r_s}. The inter-vehicular
distances are presented in Fig.~\ref{r_e}.  The combination of Figs.~\ref{r_p}-\ref{r_e} shows that
the platoon well tracks the planning trajectory to avoid the red light, in the meanwhile, maintains the desired distances between the vehicles to ensure safety.

\begin{figure}[!h]
  \centering
  \includegraphics[width=0.7\linewidth]{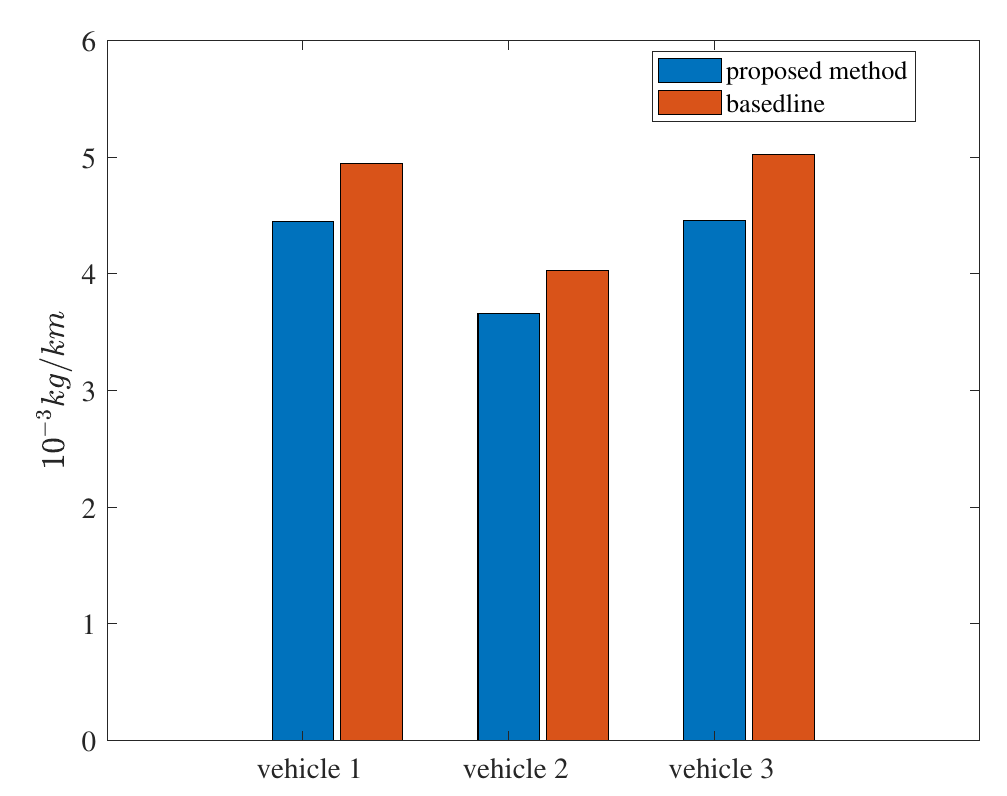}\\
  \caption{The fuel consumption.}
\label{energy}
\end{figure}
\subsection{Fuel Consumption Comparison}
For the platoon fuel consumption,
we compare our method to the standard adaptive cruise control (ACC) method viewed as the benchmark.
For the ACC method described in \cite{chen2018real}, the leader vehicle tries to maintain the speed in the desired value (cruise) and the follower follows its nearest preceding vehicle and only considers safe distance. The ACC method does not consider the traffic light information, and we know that using the traffic light information to avoid red lights can  significantly reduce the fuel consumption \cite{homchaudhuri2016fast}. Thus, to make this a fair comparison, we assume that the platoon travels on a road without traffic lights (i.e., $p_{j}=+\infty$).
The comparative result about the energy consumption is given in Fig.~\ref{energy}, which shows that using the proposed method, the platoon consumes less energy compared to the standard ACC.

\subsection{Comparison of the Effect of Avoiding Red Light}
The effects of avoiding red light are compared via simulation in this subsection between the soft constraint method \cite{homchaudhuri2016fast}  and the proposed method of this paper.
\begin{figure}[!h]
  \centering
  \includegraphics[width=0.7\linewidth]{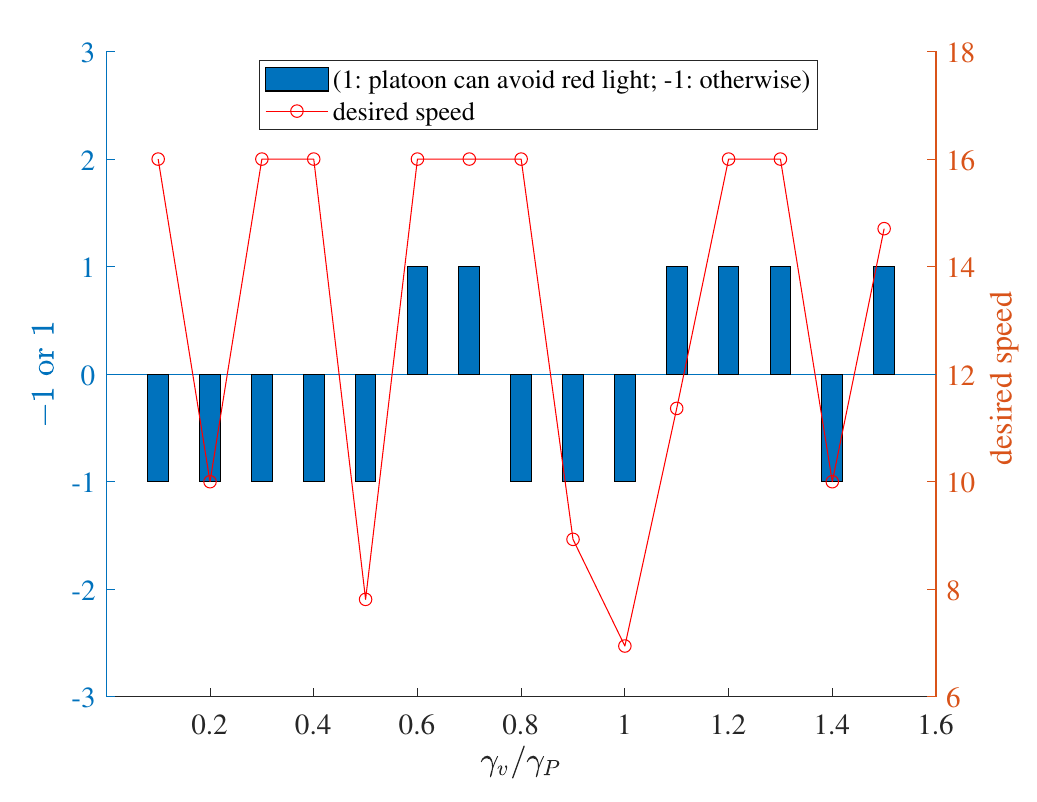}\\
  \caption{The outcome of avoiding red light using soft constraints for long-term planning.}
\label{soft}
\end{figure}

\begin{figure}[!h]
  \centering
  \includegraphics[width=0.7\linewidth]{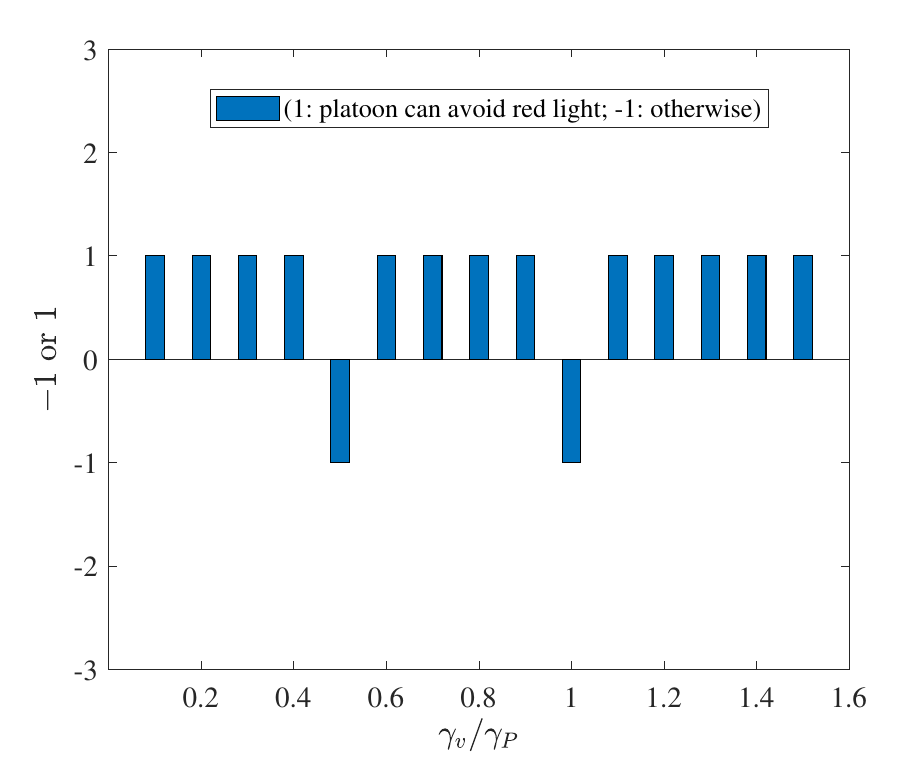}\\
  \caption{The outcome of avoiding red light using our method for long-term planning.}
\label{strict}
\end{figure}

For the soft constraint method, the speed is adjusted through a penalty term in the cost function (i.e., $\gamma_{v}(v_{1}(k)-v_{ref}(k))^{2}$), which is called the soft constraint.
The desired speed $v_{ref}(k)$ is chosen based on the traffic light signal.
The green light
durations $t_{j}^{r}$ is randomly generated in the range $[7,~22]$ by  MATLAB command $\textmd{randi}([7,~22])$. Following the model in \cite{homchaudhuri2016fast}, the desired speed $v_{ref}(k)$ is chosen to be $\bar{\psi}^{[j]}$.
The weight parameters are set to be $\gamma_{P}=0.4$, $\gamma_{B}=0.01$, $\gamma_{s}=0.001$
The setting of the other parameters is identical to the setup described in subsection \ref{ltp}, except for the $\gamma_{v}$, which is a variable for this comparison simulation (see the x-coordinate of Figs.~\ref{soft}--\ref{strict}).
The outcomes of avoiding red light w.r.t. the parameter $\gamma_{v}/\gamma_{P}$ are simulated and presented in Fig.~\ref{soft}, where $\gamma_{P}$ is the weight of the fuel consumption term in the cost function. The simulation shows that the platoon can well avoid the red lights by the soft constraint only when  $\gamma_{v}/\gamma_{P}$ is large. It is known that the energy efficiency decreases as the coefficient  $\gamma_{v}/\gamma_{P}$ increases. How to choose a proper value of $\gamma_{v}/\gamma_{P}$ is a challenging task that requires further research.

%
The simulation outcome of the strict constraint method is shown in Fig.~\ref{strict}, which shows that the platoon can avoid red light in most circumstances even if $\gamma_{v}/\gamma_{P}$ is a small value. This means that using the proposed method, the platoon can avoid red light with a high probability, and in the meanwhile, the energy efficiency can be further improved by assigning a smaller value for $\gamma_{v}/\gamma_{P}$.
Comparing Fig.~\ref{soft} with Fig.~\ref{strict}, implies that the strict constraint method proposed in this paper is more effective for the platoon avoiding the red lights.


\section{conclusion}
This paper focuses on the control algorithm design for the vehicle platoon to avoid red light and improve the fuel efficiency. A two-layer algorithm framework is proposed.
At the first layer, the long-term motion planning model is defined based on the reachability analysis of the platoon to the green light windows. An event-triggered mechanism is proposed to operate the long-term model.
At the second layer, the long-term motion planning data is used as input to design the platoon real time controllers.
A Newton's method-based algorithm is implemented to effectively solve both the long-term planning and real time control problems.

\bibliographystyle{IEEEtran}
\bibliography{bibfile}
\end{document}